\documentclass[journal]{IEEEtran}
\usepackage{subcaption}
\usepackage[dvips]{graphicx}
\usepackage{url}
\usepackage{xcolor}
\usepackage{amsmath, amssymb, amsfonts}
\usepackage{multirow}
\usepackage[hidelinks]{hyperref}
\usepackage{graphicx}
\usepackage{dsfont}
\usepackage{tikz}
\usepackage{pgfplots}
\usepgfplotslibrary{groupplots}
\pgfplotsset{compat=1.17} 

\newcommand{\onehot}{\operatorname{one\_hot}}
\DeclareMathOperator*{\argmax}{\arg\max}

\definecolor{tueScharlaken}{HTML}{C81919}
\definecolor{c1}{RGB}{31,119,180} 
\definecolor{c2}{RGB}{255,127,14}  

\begin{document}

\title{Categorical Unsupervised Variational Acoustic Clustering}


\author{Luan Vin\'icius Fiorio, Ivana Nikoloska, and Ronald M. Aarts
\thanks{This work was supported by the Robust AI for SafE (radar) signal processing (RAISE) collaboration framework between Eindhoven University of Technology and NXP Semiconductors, including a Privaat-Publieke Samenwerkingen-toeslag (PPS) supplement from the Dutch Ministry of Economic Affairs and Climate Policy.}
\thanks{L. V. Fiorio, I. Nikoloska, and R. M. Aarts are with the Eindhoven University of Technology, Eindhoven, 5612 AP, The Netherlands (e-mails: l.v.fiorio@tue.nl, i.nikoloska@tue.nl, r.m.aarts@tue.nl).}}

\markboth{PREPRINT}
{Shell \MakeLowercase{\textit{et al.}}: Bare Demo of IEEEtran.cls for IEEE Journals}
\maketitle

\begin{abstract}
We propose a categorical approach for unsupervised variational acoustic clustering of audio data in the time-frequency domain. The consideration of a categorical distribution enforces sharper clustering even when data points strongly overlap in time and frequency, which is the case for most datasets of urban acoustic scenes. To this end, we use a Gumbel-Softmax distribution as a soft approximation to the categorical distribution, allowing for training via backpropagation. In this settings, the softmax temperature serves as the main mechanism to tune clustering performance. The results show that the proposed model can obtain impressive clustering performance for all considered datasets, even when data points strongly overlap in time and frequency.
\end{abstract}


\IEEEpeerreviewmaketitle


\section{Introduction}

\IEEEPARstart{T}{he} use of classification and clustering algorithms that allow for specialized processing are crucial for hardware-constrained applications like hearing aids \cite{yook2015environmentadaptive, lamarche2010adaptive}. However, clustering algorithms require labels which can be scarce, as is the case for urban acoustic scenes datasets \cite{fiorio2023semisupervised, han2016semisupervised}, where they usually represent abstractions, e.g., the place where each audio file was recorded \cite{mesaros2018tau}. To optimize specialized processing, we need an unsupervised clustering method based on relevant characteristics of the acoustic signal \cite{fiorio2025unsupervised} that is able to process complex and overlapped audio data.

While traditional methods struggle to cluster high-dimensional audio signals \cite{foote1999overview}, variational autoencoders are a feasible option as they can learn without the need for labels \cite{kingma2014autoencoding} and generally do not require a massive number of parameters, as observed for other generative approaches \cite{kaplan2020scalinglawsneurallanguage}. Within the scope of interest, variational autoencoders were previously applied for unsupervised image clustering \cite{ugur2020variational, jiang2017variational, dilokthanakul2017deep}, and recently for the unsupervised clustering of spoken digits \cite{fiorio2025unsupervised}.

We build upon \cite{fiorio2025unsupervised} by modifying the unsupervised variational acoustic clustering (UVAC) model to a categorical UVAC, inspired by the generative semi-supervised model \cite{kingma2014semisupervised}. The proposed approach employs a categorical distribution, providing efficient clustering performance even for datasets with strongly overlapped data points, as urban acoustic scenes. This is done using a Gumbel-Softmax (GS) distribution \cite{jang2017categorical} for categorical reparametrization, allowing for training via backpropagation, with the softmax temperature serving as a mechanism to tune clustering performance. The results show impressive unsupervised clustering performance for a spoken digits dataset -- considered for comparison with the baseline \cite{fiorio2025unsupervised} -- and for two real-world datasets of urban acoustic scenes.

\section{Variational inference}

We consider a dataset $\mathbf{X} = \{\mathbf{x}^{(i)}\}^{N}_{i=1}$ with $N$ independent and identically distributed (i.i.d.) samples. Each observation $\mathbf{x}^{(i)}$ is considered to be generated by a class $k$ of a categorical latent variable $y$ and a continuous latent variable $\mathbf{z}$, similarly to the M2 model in \cite{kingma2014semisupervised}. The joint distribution is given by
\begin{equation}
    p_\theta(\mathbf{x^{(i)}}, y, \mathbf{z}) = p_\theta(\mathbf{x^{(i)}}|y,\mathbf{z}) p_\theta(y) p_\theta(\mathbf{z}),
\label{eq:generative_model}
\end{equation}
with model parameters $\theta$, and $y$ and $\mathbf{z}$ independent. We are interested in the true posterior $p_\theta(y, \mathbf{z} | \mathbf{x}^{(i)})$, as it tells us how likely each latent configuration is given an observed data point. From Bayes theorem, we can use the marginal likelihood 
\begin{equation}
    p_\theta(\mathbf{x}^{(i)}) = \sum_{y} \int p_\theta(\mathbf{x^{(i)}}, y, \mathbf{z}) d\mathbf{z}
\label{eq:marginalization}    
\end{equation}
to obtain the true posterior, with parameters $\theta$ that can be obtained by maximizing the log-likelihood $\log p_\theta(\mathbf{x}^{(i)})$. However, \eqref{eq:marginalization} has no closed-form solution for most real-world problems \cite{kingma2014autoencoding}. 

\begin{figure}[!t]
    \centering
    \includegraphics[width=1.0\linewidth]{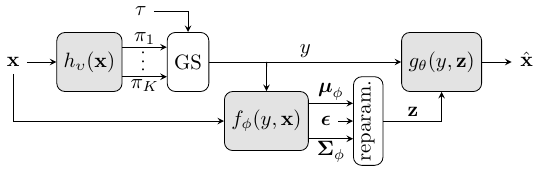}
    \caption{Diagram of the categorical UVAC model. $q_\upsilon(y|\mathbf{x})$ is represented by a NN $h_\upsilon(\mathbf{x})=[\boldsymbol{\pi}_k]$ and $\tau$ is the Gumbel-Softmax temperature. $f_\phi(y,\mathbf{x})=[\boldsymbol{\mu}_\phi, \boldsymbol{\Sigma}_\phi]$ is a NN equivalent to $q_\phi(\mathbf{z}|y,\mathbf{x})$. The decoder $p_\theta(\mathbf{x}|y,\mathbf{z})$ is a NN $g_\theta(y,\mathbf{z}) = [\hat{\mathbf{x}}]$.}
    \label{fig:M2_model}
    \vspace{-5mm}
\end{figure}

To solve the intractability of \eqref{eq:marginalization}, we define a \emph{variational} distribution with parameters $\upsilon$ and $\phi$,
\begin{equation}
    q_{\upsilon,\phi}(y, \mathbf{z} | \mathbf{x}^{(i)}) = q_\phi(\mathbf{z}|\mathbf{x}^{(i)}, y) q_\upsilon(y|\mathbf{x}^{(i)}),
\label{eq:inference_model}
\end{equation}
such that $ q_{\upsilon,\phi}(y, \mathbf{z} | \mathbf{x}^{(i)}) \approx p_\theta(y, \mathbf{z} | \mathbf{x}^{(i)})$. Moreover, we rewrite \eqref{eq:marginalization} using \eqref{eq:inference_model} as
\begin{equation}
    \log p_\theta(\mathbf{x}^{(i)}) = \log \sum_{y} \int  q_{\upsilon, \phi}(y, \mathbf{z} | \mathbf{x}^{(i)}) \frac{p_\theta(\mathbf{x^{(i)}}, y, \mathbf{z})}{q_{\upsilon,\phi}(y, \mathbf{z} | \mathbf{x}^{(i)})} d\mathbf{z},
\label{eq:variational_marginalization}    
\end{equation}
where $\sum_{y} \int q_{\upsilon,\phi}(y, \mathbf{z} | \mathbf{x}^{(i)}) [\cdot] d\mathbf{z}$ represents the expectation over $q_{\upsilon,\phi}(y, \mathbf{z} | \mathbf{x}^{(i)})$ applied to $[\cdot]$, i.e., $\mathbb{E}_{q_{\upsilon,\phi}(y, \mathbf{z} | \mathbf{x}^{(i)})}[\cdot]$. Rewriting \eqref{eq:variational_marginalization} with the expectation operator gives us
\begin{equation}
    \log p_\theta(\mathbf{x}^{(i)}) = \log \mathbb{E}_{q_{\upsilon,\phi}(y, \mathbf{z} | \mathbf{x}^{(i)})} \left[ \frac{p_\theta(\mathbf{x^{(i)}}, y, \mathbf{z})}{q_{\upsilon,\phi}(y, \mathbf{z} | \mathbf{x}^{(i)})} \right].
\label{eq:expectation_marginalization}    
\end{equation}
We can apply Jensen's inequality to \eqref{eq:expectation_marginalization}, which yields
\begin{equation}
    \log p_\theta(\mathbf{x}^{(i)}) \geq \mathbb{E}_{q_{\upsilon,\phi}(y, \mathbf{z} | \mathbf{x}^{(i)})} \left[ \log \frac{p_\theta(\mathbf{x^{(i)}}, y, \mathbf{z})}{q_{\upsilon,\phi}(y, \mathbf{z} | \mathbf{x}^{(i)})} \right],
\end{equation}
whereby, by expanding the rightmost part with \eqref{eq:generative_model} and \eqref{eq:inference_model}, we can define the variational lower bound $\mathcal{L}^{(i)}(\theta, \upsilon,\phi)$ as
\begin{multline}
    \mathcal{L}^{(i)}(\theta, \upsilon, \phi) = \mathbb{E}_{q_{\upsilon,\phi}(y, \mathbf{z} | \mathbf{x}^{(i)})}[\log  p_\theta(\mathbf{x^{(i)}}|y,\mathbf{z})] \\
    - \lambda(D_{KL}(q_\phi(\mathbf{z}|y, \mathbf{x}^{(i)}) || p_\theta(\mathbf{z})) + D_{KL}(q_\upsilon(y|\mathbf{x}^{(i)})||p_\theta(y))).
\label{eq:elbo}    
\end{multline}
In \eqref{eq:elbo}, $D_{KL}(\cdot||\cdot)$ is the Kullback-Leibler (KL) divergence between two distributions and $\mathbb{E}_{q_{\upsilon,\phi}(y, \mathbf{z} | \mathbf{x}^{(i)})}[\log p_\theta(\mathbf{x^{(i)}}|y,\mathbf{z})]$ is the reconstruction error. Additionally, we include a non-trainable Lagrangian variable $\lambda$ for smoother training, which should be tuned based on data. The complete model with generation \eqref{eq:generative_model} and inference \eqref{eq:inference_model} is shown in Figure~\ref{fig:M2_model}. In this model, each encoder and decoder are represented as a neural network (NN). To sample from a categorical distribution, the model uses a Gumbel-Softmax distribution \cite{jang2017categorical}, which we detail in the following. 

\subsection{Sampling from a categorical distribution}

\begin{figure}[!t]
    \centering
    \begin{tikzpicture}
    \hspace{-3mm} 
      \begin{groupplot}[
          group style={
            group size=2 by 2,
            horizontal sep=1.75cm,
            vertical sep=1.4cm,
          },
          scale=0.7,
          width=4.75cm,
          height=4cm,
          xlabel={$k$},
          xtick={1,10},
          ylabel={$y$},
          xlabel style={yshift=1.5mm},
          title style={yshift=-1.5mm},
          ybar,
          grid=both,
          grid style={line width=.1pt, draw=gray!20},
          major grid style={line width=.2pt,draw=gray!50},
      ]
      
      \nextgroupplot[
          title={\textcolor{black}{$\tau = 0.01$}},
          ymin=0, ymax=1.1,
      ]
      \addplot[fill=black!50!white, bar width=6pt] coordinates {
          (1,0.0000) (2,0.0000) (3,0.0000) (4,0.0000) (5,3.1838e-32)
          (6,1.0000) (7,1.1245e-35) (8,0.0000) (9,0.0000) (10,1.2752e-43)
      };
    
      \nextgroupplot[
          title={\textcolor{black}{$\tau = 0.5$}},
          ymin=0, ymax=0.55,
          ]
      \addplot[fill=black!50!white, bar width=6pt] coordinates {
          (1,0.0349) (2,0.0203) (3,0.0119) (4,0.0604) (5,0.1218)
          (6,0.5193) (7,0.1039) (8,0.0532) (9,0.0023) (10,0.0720)
      };
    
      \nextgroupplot[
          title={\textcolor{black}{$\tau = 1.0$}},
          ymin=0, ymax=0.30,
      ]
      \addplot[fill=black!50!white, bar width=6pt] coordinates {
          (1,0.0712) (2,0.0543) (3,0.0415) (4,0.0937) (5,0.1330)
          (6,0.2747) (7,0.1228) (8,0.0880) (9,0.0185) (10,0.1023)
      };
    
      \nextgroupplot[
          title={\textcolor{black}{$\tau = 100.0$}},
          ymin=0, ymax=0.11,
          ytick={0, 0.05, 0.1}, 
          yticklabels={0, 0.05, 0.1},
      ]
      \addplot[fill=black!50!white, bar width=6pt] coordinates {
          (1,0.0999) (2,0.0996) (3,0.0993) (4,0.1002) (5,0.1005)
          (6,0.1012) (7,0.1004) (8,0.1001) (9,0.0985) (10,0.1002)
      };
    
      \end{groupplot}
    \end{tikzpicture}
    \vspace{-2mm}
    \caption{Gumbel-Softmax distribution example plot for 10 classes with different values of $\tau$.}
    \label{fig:gumbel_plot}
    \vspace{-5mm}
\end{figure}
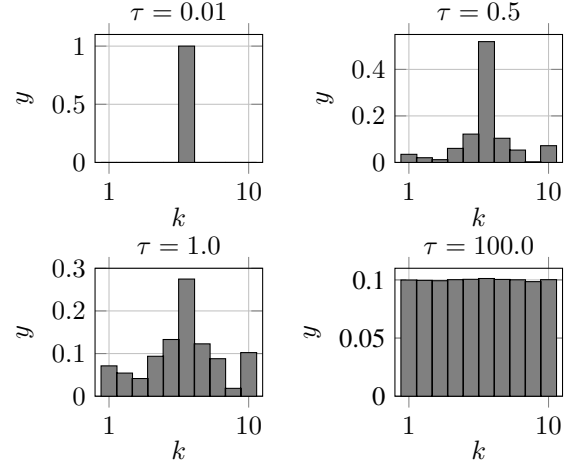

The basis of the approach comes from the Gumbel-Max trick \cite{maddison2015asampling}. Consider a categorical distribution with class probabilities $[\pi_1, \ ..., \pi_K]$, defined as
\begin{equation}
    {y}_{\mathds{1}} = \onehot \left( \argmax_k [\log \pi_k + g_k] \right) \!,
\label{eq:gumbel_max}        
\end{equation}
where $g_k \sim -\log(-\log(u))$ are i.i.d. samples drawn from the Gumbel distribution with $u \sim \mathcal{U}(0,1)$. Thereby, sampling from a discrete distribution is reduced by applying noise to a deterministic function. However, the $\argmax$ operator is not differentiable. To address this issue, we change the $\arg \max$ operator by its differentiable approximation, the softmax function, and we obtain the Gumbel-Softmax distribution \cite{jang2017categorical}:
\begin{equation}
    y = \frac{\exp \left( (\log \pi_k + g_k)/\tau \right)}{\sum_{j=1}^{K} \exp \left( (\log \pi_j + g_j)/\tau \right)}, \quad \text{for } k = 1,\text{...}, K.
\label{eq:gumbel_softmax}    
\end{equation}
The temperature $\tau$ controls the extent to which the GS approaches a categorical distribution. As $\tau \rightarrow 0$, the continuous GS \eqref{eq:gumbel_softmax} becomes the categorical Gumbel-Max distribution \eqref{eq:gumbel_max}. On the other hand, $\tau \rightarrow \infty$ makes the GS \eqref{eq:gumbel_softmax} a uniform distribution. To illustrate, Figure~\ref{fig:gumbel_plot} shows the GS distribution plot for 10 classes, with different values of $\tau$. The softmax temperature $\tau$ is specially interesting when the GS distribution is used for clustering, as a smaller $\tau$ results in more distinct and dense clusters. Therefore, as a natural choice, we choose to cluster over $y$ with a monotonic reduction of $\tau$ over training. 

\subsection{Architecture}
\label{sec:architecture}

The categorical UVAC is shown in Figure~\ref{fig:M2_model}. For the discriminative inference encoder $q_\upsilon(y|\mathbf{x}^{(i)})$, we consider a NN denoted by $h$ with parameters $\upsilon$ that outputs $K$ (soft) probabilities, $h_\upsilon(\mathbf{x}^{(i)}) = [\pi_1,\ ..., \pi_K]$. Moreover, the inference encoder $q_\phi(\mathbf{z}|y, \mathbf{x}^{(i)})$ is represented by a NN denoted by $f$ with parameters $\phi$ such that $f_\phi(y,\mathbf{x}) = [\boldsymbol{\mu}_\phi, \boldsymbol{\Sigma}_\phi]$, which outputs are used for sampling $\mathbf{z}$ via reparametrization trick \cite{kingma2014autoencoding} as $\mathbf{z} = \boldsymbol{\mu}_\phi + \boldsymbol{\Sigma}_{\phi}^{1/2} \boldsymbol{\epsilon}$, with an auxiliary random variable $\boldsymbol{\epsilon} \sim \mathcal{N}(\mathbf{0}, \mathbf{I})$. The generative decoder is given by a NN denoted by $g$ with parameters $\theta$ that reconstructs data $\mathbf{x}^{(i)}$, as $g_\theta(y,\mathbf{z}) = [\hat{\mathbf{x}}^{(i)}]$. The NNs follow a similar architecture to that of \cite{fiorio2025unsupervised}, a state-of-the-art convolutional-recurrent autoencoder approach \cite{shaug2024speechenhancement, zhao2018convolutionalrecurrent}, described as follows.

Both inference networks $h$ and $f$ are composed by four 2D convolutional layers followed by two gated recurrent unit (GRU) layers. The convolutions have output channels 16, 32, 64, and 128, while both GRU layers have 128 output nodes. Every convolutional layer is followed by 2D batch normalization and a ReLU activation function. After the last GRU layer, model $h$ contains a linear layer that outputs $K$ values, which are used as input to the GS distribution. Differently, after the last GRU layer of $f$, we use a linear layer to convert the encoder dimension to a latent dimension $d_{\mathbf{z}}$. The decoder model $g$ is composed by four 2D transposed convolutional layers with output channels 64, 32, 16, and 1, where every layer is followed by ReLU activation except the last, which uses a sigmoid function. All convolutional kernels are (8,8) with stride (2,2) and padding (3,3). Note that, whenever the input to a NN model is the concatenation of two variables, we concatenate them over the channel dimension, expanding other dimensions as needed. 

\section{Experimental evaluation}

We consider two different tasks for the validation of the proposed method. First, to compare with previous work, we target the unsupervised spoken digit recognition. Second, as a more challenging task, we devise the unsupervised clustering of urban acoustic scenes, composed of real-world background sound recordings. Both cases are described in the following.

\subsection{Tasks}

\subsubsection{Spoken digit recognition}

spoken digit recognition consists of identifying which digit was spoken in an audio utterance. The main acoustic feature is the digit itself, while other sound characteristics are minor features. For such a task, we use the AudioMNIST dataset \cite{becker2023audiomnist} with 30000 audio samples, being 24000 randomly chosen for training, 3000 for validation, and 3000 for testing. Each file contains the recording of a spoken digit in english. We use the same pre-processing of data as described in \cite{fiorio2025unsupervised}, feeding a 1-second long magnitude spectrogram with trimmed frequency bins to the NNs.

\subsubsection{Urban acoustic scene classification}
\label{ssec:urban}
we devise the classification of urban acoustic scenes using real-world datasets. Such a task is much more challenging as the (background) sound features resemble noise and greatly overlap in time and frequency. We consider two different datasets: TAU2019 \cite{mesaros2018tau}, with 1200 audio recording files of 10 seconds from different acoustic scenes; and the UrbanSound8K (US8K) \cite{salamon2014us8k} dataset, with 8732 labeled sound excerpts of 10 different acoustic scenes, mostly 4-seconds long. Differently from AudioMNIST, we now use a mel-frequency cepstrum to reduce the input dimension without limiting frequency range, as acoustic scene classification can benefit from the broader frequency range.

We resample data to 16 kHz. The US8K audio files are zero-pad to four seconds, where we also employ a voice activity detection mask for the calculation of the reconstruction error, avoiding the network to cluster zero-padding \cite{fiorio2025unsupervised}. We obtain a short-term Fourier transform (STFT) of 960 samples with a Hann window the same size and 50\% overlap. Lastly, we apply a mel-frequency scale with 128 bins. The cepstrum is normalized by its mean and variance, and a min-max normalization to limit values from 0 to 1. We feed the NNs with a time-context window \cite{fiorio2024spectral} of 4 seconds for the US8K dataset and 10 seconds for the TAU2019 dataset, which are the maximum duration of the dataset's files.

Differently from AudioMNIST case, urban scene classification has direct application for hearing aids \cite{yook2015environmentadaptive, lamarche2010adaptive}, where different acoustic scenes result in different processing, which is proportional to the constraints of the device. We take two cases into account: the same number of clusters as labels in the dataset; and a reduced number of clusters. Specifically for the considered urban acoustic scene datasets, we consider 10 clusters for the higher end, as it matches the number of labels in the dataset. For the lower end, we take 5 clusters into account, as it is a significant reduction from 10, merging similar clusters, but still sufficient for effectively calculating clustering metrics. In practice, a higher number of clusters results in a more complex and ``specialized'' processing, representing a higher-end version of a hearing aid device. On the other hand, the lower cluster number could represent a more affordable version of the same device. 

For this task, we cannot expect high (unsupervised) accuracy from the networks since the audio clips resemble background noise, e.g., TAU2019 have very abstract labels as they tell us where the recordings were made -- ``airport'', ``metro station'', ``tram''. Notice that, even when we listen to the TAU2019 audio files, it is difficult to tell which one belongs to each scene. With unsupervised clustering, we separate files based on their main features and statistical behavior, which present a significant benefit for specialized processing when compared to directly using labels such as recording location. 

\begin{table*}[!t]
    \centering
    \caption{Clustering metrics on the test set of the mentioned datasets, either considering labels as clusters or by applying K-means, UVAC and categorical (Cat.) UVAC as clustering methods, averaged over 10 independent runs. *results from \cite{fiorio2025unsupervised}.}
    \begin{tabular}{c c c c c c c c}
    \hline
        \textbf{Dataset} & \textbf{Clusters} & \textbf{Method} & \textbf{Accuracy} (\%) \textcolor{darkgray}{$\uparrow$} & \textbf{NMI} \textcolor{darkgray}{$\uparrow$} & \textbf{Silhouette} \textcolor{darkgray}{$\uparrow$} & \textbf{DBI} \textcolor{darkgray}{$\downarrow$} & \textbf{CHI} $\times10^{3}$ \textcolor{darkgray}{$\uparrow$} \\
    \hline        
        \multirow{4}{*}{AudioMNIST} & \multirow{4}{*}{10} & \textcolor{gray}{None (labels)}* & \textcolor{gray}{100.00} & \textcolor{gray}{1.00} & \textcolor{gray}{-0.04} & \textcolor{gray}{5.56} & \textcolor{gray}{0.10} \\
        &  & K-means* & 18.40 & 0.10 & 0.13 & 2.04 & 0.69 \\
        &  & UVAC* & 70.78 & 0.71 & 0.21 & 1.61 & 0.54 \\
        &  & Cat. UVAC ($\lambda=0.5$) & 76.30 & 0.78 & \textbf{\textcolor{black}{0.97}} & \textbf{\textcolor{black}{0.07}} & \textbf{\textcolor{black}{40.14}} \\
    \hline
        \multirow{7}{*}{TAU2019} & \multirow{4}{*}{10} & \textcolor{gray}{None (labels)} & \textcolor{gray}{100.00} & \textcolor{gray}{1.00} & \textcolor{gray}{-0.05} & \textcolor{gray}{9.17} & \textcolor{gray}{0.11} \\
        &  & K-means & 25.20 & 0.19 & 0.03 & 3.48 & 0.28 \\
        &  & UVAC & 28.73 & 0.17 & -0.02 & 5.07 & 0.25 \\
        &  & Cat. UVAC ($\lambda=2.0$) & 23.63 & 0.15 & \textbf{\textcolor{black}{0.77}} & \textbf{\textcolor{black}{0.33}} & \textbf{\textcolor{black}{6.97}} \\   
    \cline{2-8}
        & \multirow{3}{*}{5} & K-means & -- & -- & 0.06 & 3.12 & 0.53 \\        
        &  & UVAC & -- & -- & 0.02 & 3.41 & 0.34 \\
        &  & Cat. UVAC ($\lambda=2.0$) & -- & -- & \textbf{\textcolor{black}{0.79}} & \textbf{\textcolor{black}{0.30}} & \textbf{\textcolor{black}{14.76}} \\  
    \hline
        \multirow{7}{*}{UrbanSound8K} & \multirow{4}{*}{10} & \textcolor{gray}{None (labels)} & \textcolor{gray}{100.00} & \textcolor{gray}{1.00} & \textcolor{gray}{-0.06} & \textcolor{gray}{4.96} & \textcolor{gray}{0.04} \\
        &  & K-means & 34.17 & 0.30 & 0.14 & 1.90 & 0.20 \\
        &  & UVAC & 35.46 & 0.30 & 0.09 & 2.35 & 0.11 \\
        &  & Cat. UVAC ($\lambda=2.0$) & 25.52 & 0.15 & \textbf{\textcolor{black}{0.73}} & \textbf{\textcolor{black}{0.39}} & \textbf{\textcolor{black}{1.17}} \\   
    \cline{2-8}
        & \multirow{3}{*}{5} & K-means & -- & -- & 0.18 & 1.63 & 0.32 \\
        &  & UVAC & -- & -- & 0.10 & 2.33 & 0.14 \\
        &  & Cat. UVAC ($\lambda=2.0$) & -- & -- & \textbf{\textcolor{black}{0.78}} & \textbf{\textcolor{black}{0.32}} & \textbf{\textcolor{black}{3.19}} \\         
    \hline
    \end{tabular}
    \label{tab:results}
    \vspace{-4mm}
\end{table*}

\subsection{Metrics}

Five metrics are considered: i) Unsupervised accuracy -- the Hungarian algorithm \cite{kuhn1955hungarian} is used to match cluster labels to truth labels. The matched labels are used for calculating accuracy, ranging from 0 to 100\%; ii) Normalized mutual information (NMI) -- evaluates how much information is shared between cluster and truth labels \cite{vinh2010information}, ranging from 0 to 1, being proportional to accuracy; iii) Silhouette score -- measures how similar a data point is to its cluster in comparison to different clusters \cite{rousseeuw1987silhouettes}. The value reflects on cohesion -- how similar data points are in a cluster -- and separation -- how separate the clusters are. It ranges from -1 (bad) to 1 (good); iv) Davies-Bouldin index (DBI) -- the average similarity rate of each cluster with its most similar cluster \cite{davies1979cluster}, being an indication of compactness and separation of clusters. The DBI score ranges from 0 (good) to infinity (bad); v) Cali\'nski-Harabasz index (CHI) -- measures the ratio of the sum of between-cluster to within-cluster dispersion \cite{calinski1974dendrite} -- distinctiveness and compactness. It ranges from 0 (bad) to infinity (good). The CHI is added on top of the metrics considered in \cite{fiorio2025unsupervised} for a broader analysis. Accuracy and NMI tell us how predicted clusters deviate from labels, and the others show how good the clusters are.

\subsection{Baselines}

As a main baseline, we consider the UVAC model from \cite{fiorio2025unsupervised}. We apply it with the exact same architecture as defined in the original paper. We expect UVAC to have a good clustering performance for the better-defined dataset, AudioMNIST, as the main feature -- the spoken digit -- is free of noise and other overlapping audio content. However, since UVAC takes a Gaussian mixture model (GMM) for clustering, it should fail for noise-resembling data, as the GMM lacks a mechanism to enforce clustering behavior, like the temperature $\tau$ of the GS. Additionally, we consider a classical approach, the K-means algorithm \cite{hartigan1979kmeans}, as a low-complexity baseline.

\subsection{Hyperparameters}

We train the model described in Section~\ref{sec:architecture} for 500 epochs, separately for each considered dataset. All NN parameters $\theta$, $\upsilon$, and $\phi$ are jointly optimized by maximizing \eqref{eq:elbo} with the Adam algorithm \cite{kingma2015adam}. For all cases of the proposed categorical UVAC and the baseline UVAC model \cite{fiorio2025unsupervised}, the initial learning rate is of $5\times10^{-4}$, with exponential decay until the last epoch, reaching $5\times10^{-5}$. The Gumbel-Softmax temperature $\tau$ in \eqref{eq:gumbel_softmax} of the categorical UVAC is annealed from 1.0 to 0.5 over the epochs, providing a ``more uniform'' distribution at the beginning of training, avoiding local minima, and converging to an ``almost categorical'' representation at end of training, enforcing clustering behavior. The value of $\lambda$ in \eqref{eq:elbo} is chosen experimentally and we try to keep it as low as possible to avoid disturbing the balance of the variational lower bound, while still avoiding the collapse -- unintended merging -- of clusters. For AudioMNIST, we use $\tau = 0.5$, while we change it to $\tau=2.0$ for both TAU2019 and UrbanSound8K datasets.

\subsection{Results}

All results are shown in Table~\ref{tab:results}. For the task of spoken digit recognition, the categorical UVAC drastically increases all unsupervised clustering metrics. The Silhouette score, importantly, is almost maxed out, indicating separate and compact clusters. The DBI and CHI indexes also achieve impressive values, reinforcing the hypothesis that the categorical distribution enforces distinct clustering behavior. Specially for AudioMNIST, we also see that the proposed method achieved high accuracy and NMI. Here, the high accuracy and NMI of spoken digits may serve as a ``sanity check'', indicating that the formed clusters are meaningful. Notice that such behavior is not expected with urban sound datasets since the files resemble background noise. The UVAC method performs the second best, where we can notice similar accuracy and NMI to its categorical version but reduced clustering metrics, showing the limitations of a Gaussian mixture model for clustering. K-means achieves insufficient metrics, only enhancing clustering when compared to the truth labels.

Moreover, the categorical UVAC achieves very high unsupervised clustering metrics for TAU2019 when 10 clusters are considered. As the number of clusters represent the same number as the classes in the dataset, we also investigate the achieved accuracy and NMI. Differently from spoken digit recognition, the urban acoustic scenes strongly overlap in time and frequency, with labels representing an abstraction, indicating where the audio files were recorded. As we can see, the best clustering metrics are achieved with the lowest accuracy and NMI, reinforcing the hypothesis from Section~\ref{ssec:urban} -- that we cannot obtain high accuracy/NMI and high clustering metrics simultaneously for such datasets. On another hand, the baseline UVAC model is not able to cluster such complex data. This shows that the GMM is insufficient for clustering data that strongly overlaps in time and frequency. K-means achieves similar behavior, failing to cluster as the method is too simple and cannot perform well with most audio data. From the clustering metrics evaluated at the labels, we see that the original classification does not form good clusters, as expected. When the number of cluster is reduced to 5, all methods improve in terms of clustering performance as nearby clusters are merged together. The categorical UVAC still performs better, while the baseline UVAC and K-means are insufficient.

Finally, the UrbanSound8K dataset also presents classes with significant overlap. This is observed once again in the reduced accuracy and NMI but high clustering metrics achieved by the categorical UVAC, showing that even though classes are better defined, their content strongly overlaps. This type of behavior confirms our aforementioned hypothesis for background-noise-like datasets. For all cases, while the baseline UVAC and K-means cannot achieve sufficient clustering, the categorical UVAC achieves high clustering performance, which metrics are increased when the number of clusters is reduced from 10 to 5. In addition, we noticed that clustering performance is dependent on the value of $\tau$. The relation between softmax temperature and clustering could be explored further, being out of scope for this paper.

\section{Conclusion}

We proposed a categorical extension for unsupervised variational acoustic clustering, which clusters audio data in the time-frequency domain. For smooth training, we employed the Gumbel-Softmax distribution, serving as a categorical approximation while still allowing for training via backpropagation. We showed out how the softmax temperature serves as a means to enhance cluster separation and compactness. The proposed approach excels in all considered cases, always achieving very high clustering metrics. As a future work, it would be useful to test the practical applications of categorical UVAC, for example in the speech enhancement of hearing aids.


\section{Acknowledgment}

The authors would like to thank the insightful comments and revision of Bruno Defraene, Johan David, Alex Young, Yan Wu, Frans Widdershoven, and Wim van Houtum.

\clearpage
\bibliographystyle{IEEEtran}
\bibliography{refs}

\end{document}